\documentclass[10pt,twocolumn]{article}

\usepackage{natbib}
\usepackage{txfonts}
\usepackage{balance}
\usepackage{graphicx}
\usepackage[squaren]{SIunits}
\usepackage{isolatin1}
\usepackage[unicode]{hyperref}

\makeatletter
\renewcommand\hyper@natlinkbreak[2]{#1}
\makeatother
 
\begin{document}

\def\dd{\rm{d }}
\def\Aeff{A\rm_{eff }}
\def\AQED{A\rm_{QED }}

\newcommand{\gfrac}[2]{\displaystyle\frac{#1}{#2}}

\title{Performance of the MeV gamma-ray telescopes and polarimeters of the future.
\\
$\gamma \to e^+ e^-$ in silicon-detector active targets}

\author{D.\,Bernard}

\date{}

\twocolumn[
\begin{@twocolumnfalse}
\maketitle

\vspace{-3em}

\begin{center}
LLR, Ecole Polytechnique, CNRS/IN2P3, 91128 Palaiseau, France

~

\href{https://www.astro.unige.ch/integral2019/} 
 {12th INTEGRAL conference and 1st AHEAD Gamma-ray workshop}
\\
 11 - 15 Feb. 2019, Geneva

\end{center}

\begin{abstract}
A number of techniques are being developed to solve the
monstrous sensitivity gap that exists between the energy ranges of
good sensitivity of the Compton and of the pair telescopes and to
extend polarimetry to the MeV gamma-ray world.
I characterize the properties of an active target, detailing the
various contributions to the angular resolution, using a full
five-dimensional event generator of the Bethe-Heitler differential
cross section.
With the same tool I also examine the dilution of the polarization
asymmetry induced by multiple scattering in the conversion wafer.
\end{abstract}

{\bf keywords} gamma-ray telescope -- gamma-ray polarimeter -- pair conversion -- silicon detector -- multiple scattering -- event generator
 
~

\end{@twocolumnfalse}
]

\section{Introduction}


Facing the huge sensitivity gap between the energy ranges of the
Compton and of the pair-creation telescopes
(Fig. 1 of \citet{Schoenfelder}),
and given the absence of
polarimetry measurements of cosmic sources with pair conversion,
i.e. in the MeV energy range, a number of techniques are put forward
to design high-performance MeV $\gamma$-ray telescopes and polarimeters:
(1) gas detectors (HARPO \citep{Gros:2017wyj,Bernard:2018jql});
(2) emulsion detectors (GRAINE \citep{Takahashi:2015jza,Ozaki:2016gvw});
(3) silicon detectors
(e-ASTROGAM \citep{DeAngelis:2017gra}
and
AMEGO \citep{Moiseev:2017mxg}).
These telescopes are based on the use of active targets, that is,
a scheme in which the photon converts and the two leptons
(the $e^+$ and the $e^-$) are tracked in the same detector.

I focus here on silicon detectors, providing simple analytical
expressions that describe the various contributions to the angular
resolution and examining the dilution of the polarization asymmetry
due to multiple scattering in the conversion wafer.

\section{Angular resolution}

\begin{figure}[t!]
\resizebox{\hsize}{!}{\includegraphics[clip=true]{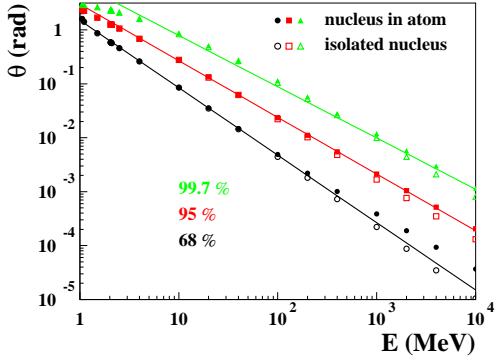}}
\caption{\footnotesize
Variation of the 68\%, 95\% and 99.7\% containment values of the angular resolution due to the missing nucleus-recoil momentum, as a function of incident photon energy $E$ (adapted from \citet{Gros:2016dmp}).
}
\label{fig:recul}
\end{figure}

Upon the pair-conversion of an MeV $\gamma$ ray in the field of a
nucleus, the final state consists of three particles, the $e^+ e^-$
pair and the ``recoiling'' nucleus, to which a small
($\kilo\electronvolt/c$ to $\mega\electronvolt/c$) momentum is transferred;
the recoiling nucleus is therefore slow and its track length is too short
to allow a measurement.
The missing recoil momentum, $q$, contributes to the angular
resolution.
As the $q$-spectrum shows a tail to high values (Fig. 3 of
\citet{Bernard:2012uf}), ``containment values'' at a given fraction
are computed (\citep{Gros:2016dmp} and Fig.~\ref{fig:recul}).
The 1-$\sigma$, 2-$\sigma$ and 3-$\sigma$ containment values are found
to vary with the photon energy, $E$, approximately like
\begin{itemize}\setlength\itemsep{-0.1em}
\item 
 $\theta_{68\,\%} \approx 1.5 \,\radian ~ \left({E}/{\mega\electronvolt} \right)^{-1.25}$,
\item 
 $\theta_{95\,\%} \approx 3.0 \,\radian ~ \left({E}/{\mega\electronvolt} \right)^{-1.05}$,
\item 
 $\theta_{99.7\,\%} \approx 7.0 \,\radian ~ \left({E}/{\mega\electronvolt} \right)^{-0.95}$.
\end{itemize}

These results were obtained using an exact sampling of the
five-dimensional Bethe-Heitler differential cross section
\citep{Bernard:2013jea} that I have made available as the
G4BetheHeitler5DModel physics model of the recently deployed
10.5 Geant4 release \citep{Bernard:2018hwf}.
None of the pre-existing physics models were found to be appropriate
for the simulation of the high-performance detectors that are being
planned nor for polarization \citep{Gros:2016zst}.

MeV $\gamma$-ray conversions produce leptons
with $\mega\electronvolt/c$ momentum that undergo significant
multiple scattering in each wafer with an RMS deflection angle
 $\theta_0 \approx (p_0/p) \sqrt{e/X_0)}$,
where $p$ is the track momentum, 
$p_0 = 13.6\,\mega\electronvolt/c$ is the multiple scattering constant \citep{PDG},
$e$ is the wafer thickness through which the lepton propagates and
$X_0$ is the wafer material radiation length.
Fitting the tracks with a Kalman filter allows multiple scattering to
be taken into account together with the single-wafer spatial
resolution in an optimal way, assuming Gaussian statistics.
The (RMS) angular resolution of a segmented detector, $\sigma_{\theta}$, is
\footnote{I have corrected a misprint.} \citep{Frosini:2017ftq}
\footnotesize
\begin{equation} 
 \sigma_{\theta} =
 \gfrac{\sigma}{l}
 \sqrt{
 \frac
 {2 \, x^3 \, \left(\sqrt{4 j - x^2} + \sqrt{- 4 j -x^2} \right)}
 { \left(\sqrt{4 j - x^2} + j x \right) \left(\sqrt{- 4 j - x^2} - j x \right)}
 }
 ,
\end{equation}
\normalsize

\begin{figure}[t!]
\resizebox{\hsize}{!}{\includegraphics[clip=true]{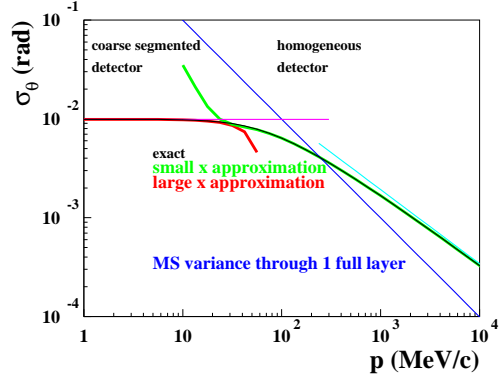}}
\caption{\footnotesize
 Single-track angular resolution of a segmented detector as a function of track momentum (black curve).
 The small-$x$ approximation, valid at high energies, is eq. (22) of \citep{Frosini:2017ftq}), green curve (the cyan line is $\sigma_{\theta} \approx \left({p}/{p_1}\right)^{-3/4}$).
 The high-$x$ approximation, valid at low energies, is eq. (25) of \citep{Frosini:2017ftq}), red curve (the magenta line is $\sigma_{\theta} \approx \sqrt{2} \sigma/l$).
}
\label{fig:tracking}
\end{figure}

where $x$ is the distance between wafers, $l$, normalized to the
detector scattering length $\lambda$
(defined in eq. (17) of \citet{Frosini:2017ftq}, see also \citet{Innes:1992ge}),
$x\equiv l/\lambda = \sqrt{({l}/{\sigma}) ({p_0}/{p}) \sqrt{{e}/{X_0}}}$,
$\sigma$ is the precision of the measurement of the track position in a wafer, 
and $j$ is the imaginary unit.
The variation of $\sigma_{\theta}$ with $p$ (Fig.~\ref{fig:tracking})
shows two regimes:
\begin{itemize}
\item at low momentum (large $x$), the coarsely-segmented detector
 approximation can be used, with $\sigma_{\theta} \approx \sqrt{2}
 \sigma/l$ (see eq. (25) of \citet{Frosini:2017ftq}).
An optimal measurement can be
 obtained simply from the position measurements in the two first
 wafers, no Kalman filter is needed.
 \item at high momentum (small $x$), the homogeneous detector approximation
 can be used, $\sigma_{\theta} \approx \left({p}/{p_1}\right)^{-3/4}$
 where $p_1$ is a momentum that characterizes the tracking-with-multiple-scattering properties of the detector, 
 $p_1 = p_0 \left({4 \sigma^2 l^4}/{X_0^3 e^3} \right)^{1/6}$
\citep{Bernard:2013jea}
(see also eq. (22) of \citet{Frosini:2017ftq}).
\end{itemize}
I used e-ASTROGAM numerical values of the detector parameters
\citep{DeAngelis:2017gra}.
These results were obtained assuming that the photon converted in the
lowest part of a wafer. In the more general case of a conversion well
within the wafer, the additional contribution of multiple scattering
in the conversion wafer must be added (for the full thickness $e$, the
blue line in Fig.~\ref{fig:tracking}).
\begin{figure}[t!]
 \resizebox{\hsize}{!}{\includegraphics[clip=true]{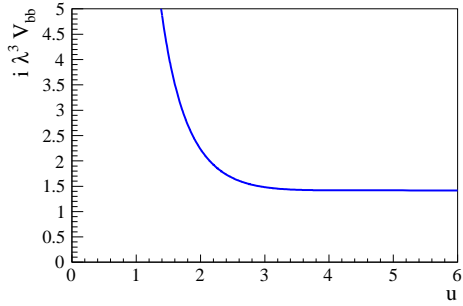}}
\caption{\footnotesize
Normalized angle variance as a function of the total detector
thickness normalized to the detector scattering length,
$u \equiv N x = N \, l / \lambda$,
 in the fully homogeneous detector approximation, $x=0$
 (several values of $x$ were examined in Fig. 4 of \citet{Frosini:2017ftq}).
}
\label{fig:u}
\end{figure}

\begin{figure}[t!]
\resizebox{\hsize}{!}{\includegraphics[clip=true]{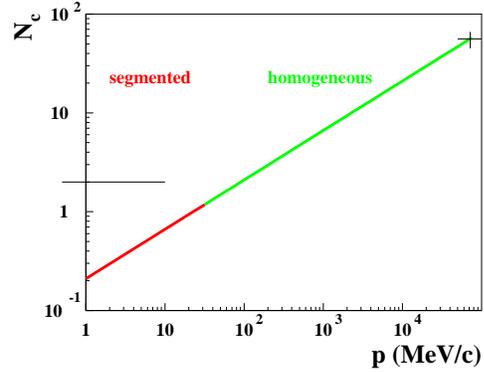}}
\caption{\footnotesize
Number of wafers that enable an optimum
 measurement, as a function of track momentum, obtained assuming a
 homogeneous-detector approximation (valid for the green part of the line). }
\label{fig:N}
\end{figure}

The above results were obtained assuming that the detector is thick enough
that the Kalman filter converged to an equilibrium: only the first
part of the track contributes significantly to the angle measurement,
after which multiple scattering floods it all.
A detector is found to be thick for a thickness of at least 2.5 
scattering lengths
(Fig.~\ref{fig:u}, and Fig. 4 of \citet{Frosini:2017ftq}),
that is for $p < 72\,\giga\electronvolt/c$ for e-ASTROGAM.
For most of the available momentum range, the optimum number of wafers
is then quite small (Fig.~\ref{fig:N}).
Fitting a longer track segment would expose the Kalman filter to
un-necessary bias induced by non-Gaussian noise contributions, such as
that from Bremsstrahlung radiation.

\begin{figure}[t!]
\resizebox{\hsize}{!}{\includegraphics[clip=true]{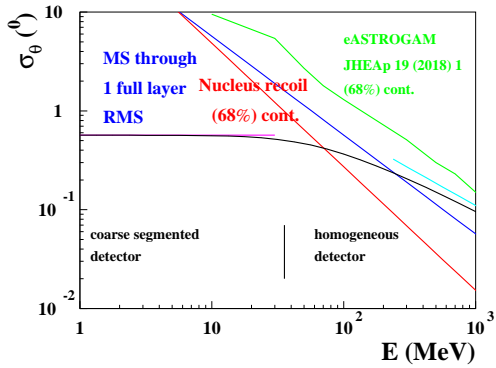}}
\caption{\footnotesize
Various contributions to the single photon angular resolution as a function of photon energy $E$.
}
\label{fig:resolution:angulaire}
\end{figure}

Assuming that the single-track angular resolution for the two leptons
translates to a single photon angular resolution with the same
expression (See Fig. 1 of \citet{Bernard:2012uf}), I obtain a global description of the angular resolution of a segmented active target telescope, Figure 
\ref{fig:resolution:angulaire}.
Significant contributions are 
\begin{itemize}\setlength\itemsep{-0.1em}
\item the missing recoil at low energy;
\item tracking and multiple scattering in the detector at high energy;
\item multiple scattering in the conversion wafer, over the whole range.
\end{itemize}

We note the agreement with the results of simulations by e-ASTROGAM
\citep{DeAngelis:2017gra} even though they most likely used a pre-10.5
Geant4 release, that is, a recoil momentum distribution different from
that of the Bethe-Heitler differential cross section
\citep{Gros:2016zst}.

\section{Polarization}

The polarimetry of high-energy photons, that is, in the energy range
for which the conversion of a photon can be recorded individually, is
performed by the analysis of the distribution of the azimuthal angle, $\phi$:
\begin{equation}
\gfrac{\dd \sigma}{\dd \phi} \propto 
\left(
1 + A \times P \cos(2(\phi - \phi_0))
\right) , 
\label{eq:modulation}
\end{equation}
where $P$ is the linear polarization fraction of the incoming radiation,
and
$A$ is the polarization asymmetry of the conversion process.
The azimuthal angle of the event measures the orientation of the final
state in the plane perpendicular to the direction of propagation of
the incoming photon.

\begin{figure}[t!]
\resizebox{\hsize}{!}{\includegraphics[clip=true]{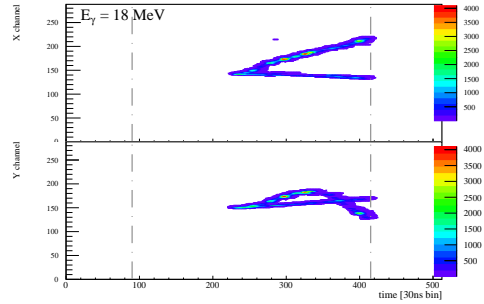}}
\caption{\footnotesize
Example of a conversion event for which the loss of azimuthal
information during the propagation of the track in the detector is
visible:
($x,t$) and ($y,t$) views of a 18\,MeV $\gamma$-ray from the BL01 beam
line at NewSUBARU (LASTI, Hyôgo U., Japan)
converting to $e^+e^-$ in the 2.1\,bar
Ar:Isobutane 95:5 gas of the HARPO TPC prototype.
For this event, track crossing is seen to take place after
$e \approx 10\,\centi\meter$ with $X_0 \approx 5600\,\centi\meter$,
that is, for $t \equiv e/X_0 \approx 1.8\,10^{-3}$. }
\label{fig:evt}
\end{figure}

Polarimetry in the pair-conversion regime with silicon detectors has
never been demonstrated, most likely due to a number of reasons,
in particular the multiple scattering undergone by the two leptons
during the propagation inside the detector
(Fig.~\ref{fig:evt})
and the difficulty of reconstructing the tracks close to the
vertex with finite strip-width detectors.

The dilution of the
polarization asymmetry due to multiple scattering in the conversion
wafer, that is, of the effective value normalized to the QED value,
$D = \Aeff / \AQED$, can be computed simply by a convolution of the differential cross section eq. (\ref{eq:modulation}) with a resolution function, yielding
$D = e^{-2 \sigma_\phi^2} $,
where $\sigma_\phi$ is the resolution of the measurement of $\phi$.
The calculation by \citep{Kotov} used the most probable value
$\hat{\theta}_{+-} = E_0 /E$, with $E_0 = 1.6\,\mega\electronvolt$
\citep{Olsen:1963zz}
of the pair opening angle $\theta_{+-}$, to obtain
 $\sigma_\phi \approx (\theta_{0,e^+}\oplus\theta_{0,e^-})/\hat{\theta}_{+-}
 \approx 24\, \radian \sqrt{e/X_0}$.
A degradation of $D$ by a factor of two is undergone for 
$\sigma_\phi =\sqrt{\ln{2}/2} \approx 0.59\,\radian$,
that is, 
after a path length of $e \approx 10^{-3} X_0$,
($e \approx 100\, \micro\meter$ for silicon or
 $e \approx 4\, \micro\meter$ for tungsten).

\begin{figure}[t!]
 \resizebox{\hsize}{!}{\includegraphics[clip=true]{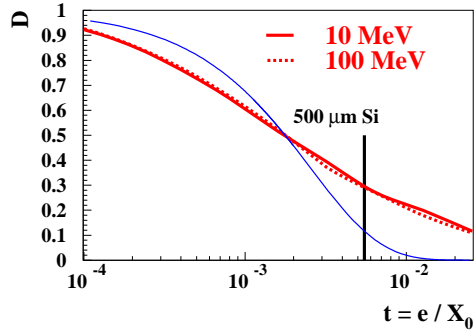}}
\caption{\footnotesize
Dilution of the polarization asymmetry, that is, the effective value
normalized to the QED value, as a function of the wafer thickness
traversed by the two leptons, normalized to the material radiation
length
(updated from \citet{Bernard:2013jea}, with the optimal definition of
the azimuthal angle, which we found to be the azimuthal angle of the
bisectrix of the azimuthal angles of the $e^-$ and of the $e^+$
\citep{Gros:2016dmp}).
The thin blue line is Kotov's $E$-independent,
$\hat{\theta}_{+-}$-based approximation.
The thick red lines are from the full (5D) simulation.
\label{fig:dilution}}
\end{figure}

The $\theta_{+-}$ distribution actually shows a tail at high values
(Fig. 7 of \citet{Bernard:2018hwf}) and there was hope that these
events would be less affected by multiple scattering.
Using the exact Bethe-Heitler event generator, I observed
\citep{Bernard:2013jea} that the dilution is found to be larger
(i.e. better) at large wafer thicknesses (hundreds of microns) than that
predicted assuming Kotov's approximation (Fig.~\ref{fig:dilution}).

\section{Conclusions, recommendations} 

For most of their energy range, the all-silicon detector telescopes
such as e-ASTROGAM and AMEGO that are being developed are thick
detectors:
\begin{itemize}\setlength\itemsep{-0.1em}
\item the tracking-with-multiple-scattering contribution to the
 resolution angle is described by the analytical expressions of
 Section 2.1.1 of \citet{Frosini:2017ftq};
\item the optimal number of wafers on which the Kalman-filter is
 applied is small and depends on photon energy; fitting longer
 track segments would make the result uselessly prone to bias induced by non
 Gaussian noise from radiation;
\item the angular resolution should not depend on the depth of the
 conversion point in the detector.
\end{itemize}

\noindent
At high energies,
\begin{itemize}\setlength\itemsep{-0.1em}
\item the homogeneous detector approximation is valid, and the use of a Kalman-filter-based tracking software is optimal.
\end{itemize}

\noindent
At low energies,
\begin{itemize}\setlength\itemsep{-0.1em}
\item the measurement of the direction of the incoming photon can be
 performed simply from the positions measured in the two first
 wafers. 
\item the impossibility to measure the momentum of the
recoiling nucleus is an important contribution to the angular
resolution, making it mandatory to use an exact simulation 
of the distribution, in angle and in magnitude of the recoil, that is
available in Geant4 release $\ge 10.5$.

Over most of the energy range, the containment angle induced by the
unmeasured recoil momentum is universal, that is, it does not depend
on the nature of the converter; only at the highest energies does the
difference in the screening of the field of the atomic electrons
induce a difference in the containment angle.
\end{itemize}

The transition between the low- and the high-energy regimes takes
place in the $1.3 < x < 2$ range, that is in, the
$35 < p < 80\,\mega\electronvolt/c$ track momentum range
for e-ASTROGAM.
The contribution of the multiple scattering in the conversion wafer is
important over the whole energy range, except for conversions in the
very bottom part of the wafer.
An event-dependent angular resolution could be considered, should the
depth of the conversion point in the wafer be estimated from the
deposited energy.

The study of the polarimetric performances of a detector necessitates
the use of an exact event generator, and of an appropriate definition
of the event azimuthal angle.
The dilution of the polarization asymmetry due to multiple
scattering in the conversion wafer is found to be much better,
that is much higher, than that calculated in the most-probable opening
angle approximation.
The dilution is also found to be independent of the photon energy.

\bibliographystyle{aa}

\begin{thebibliography}{}
\footnotesize
\bibitem[{Bernard(a 2013)}]{Bernard:2013jea} 
Bernard,~D. a,
\href{http://inspirehep.net/record/1242601}{2013, Nucl.\ Instr.\ Meth.\ A {\bf 729}, 765} 

\bibitem[{Bernard(b 2013)}]{Bernard:2012uf}
Bernard,~D. b,
\href{http://inspirehep.net/record/1198968}{2013, Nucl.\ Instr.\ Meth.\ A {\bf 701}, 225} 

\bibitem[{Bernard(a 2018)}]{Bernard:2018hwf} 
Bernard,~D. a,
\href{https://inspirehep.net/record/1657191}{2018, Nucl.\ Instr.\ Meth.\ A {\bf 899}, 85} 
 
\bibitem[{Bernard(b 2018)}]{Bernard:2018jql}
Bernard,~D. b,
\href{https://doi.org/10.1016/j.nima.2018.10.016}{Nucl.Instrum.Meth. A936 (2019) 405}


\bibitem[{De Angelis {\it et al.}(2018)}]{DeAngelis:2017gra} 
De Angelis,~A. {\it et al.},
\href{http://inspirehep.net/record/1634531}{2018, JHEAp {\bf 19}, 1}

\bibitem[{Frosini \& Bernard(2017)}]{Frosini:2017ftq} 
Frosini,~M.~\&~Bernard,~D.,~
\href{https://inspirehep.net/record/1605726}{2017, Nucl.\ Instr.\ Meth.\ A {\bf 867}, 182} 

\bibitem[{Gros {\it et al.}(2018)}]{Gros:2017wyj}
Gros,~P. {\it et al.},
\href{https://inspirehep.net/record/1606056}{2018, Astropart.\ Phys.\ {\bf 97}, 10}

\bibitem[{Gros \& Bernard(a 2017)}]{Gros:2016dmp}
Gros,~P. \& Bernard,~D. a,
\href{https://inspirehep.net/record/1498288}{2017, Astropart.\ Phys.\ {\bf 88}, 30} 

\bibitem[{Gros \& Bernard(b 2017)}]{Gros:2016zst} 
Gros,~P. \& Bernard,~D. b, 
\href{https://inspirehep.net/record/1504915}{2017, Astropart.\ Phys.\ {\bf 88}, 60} 

\bibitem[{Innes(1993)}]{Innes:1992ge}
Innes,~W.~R.,
\href{http://inspirehep.net/record/335574}{1993, Nucl.\ Instr.\ Meth.\ A {\bf 329} 238}.

\bibitem[{Kotov(1988)}]{Kotov}
Kotov,~Iu.~D.,
\href{https://link.springer.com/article/10.1007/BF00173754}{1988, Space Sci. Rev., {\bf 49}, 185}
 
\bibitem[{Moiseev {\it et al.}(2018)}]{Moiseev:2017mxg}
Moiseev,~A., {\it et al.},
\href{http://inspirehep.net/record/1686484}{2018, PoS ICRC {\bf 2017} 798}.

\bibitem[{Olsen(1963)}]{Olsen:1963zz} 
Olsen,~H., 
\href{http://inspirehep.net/record/46674}{1963, Phys.\ Rev.\ {\bf 131}, 406} 
 
\bibitem[{Ozaki {\it et al.}(2016)}]{Ozaki:2016gvw}
Ozaki,~K. {\it et al.},
\href{http://inspirehep.net/record/1455851}{2016, Nucl.\ Instr.\ Meth.\ A {\bf 833}, 165}
 
\bibitem[{Schönfelder(2004)}]{Schoenfelder}
Schönfelder,~V.,
\href{https://doi.org/10.1016/j.newar.2003.11.027}{2004, New Astron. Rev. {\bf 48}, 193}.
 
\bibitem[{Takahashi {\it et al.}(2015)}]{Takahashi:2015jza} 
Takahashi,~S. {\it et al.},
\href{http://inspirehep.net/record/1362123}{2015, PTEP 043H01}.

\bibitem[{PDG(2018)}]{PDG}
Tanabashi,~M. {\it et al.}, (Particle Data Group),
\href{http://inspirehep.net/record/1688995}{2018, Phys. Rev. D {\bf 98}, 030001}.
\end{thebibliography}

\end{document}